\documentclass[aps,prl,superscriptaddress,amsmath,amssymb,showpacs,twocolumn]{revtex4-1}
\usepackage{amssymb}
\usepackage{amsmath}
\usepackage{graphicx}
\usepackage{bm}
\usepackage{color}
\usepackage{subfigure}

\newcommand{\Q}{\dot{Q}}

\begin{document}

\title{Thermally-activated non-local amplification in quantum energy transport}

\author{Bruno Leggio}
\affiliation{Laboratoire Charles Coulomb (L2C), UMR 5221 CNRS-Universit\'{e} de Montpellier, F- 34095 Montpellier, France}

\author{Riccardo Messina}
\affiliation{Laboratoire Charles Coulomb (L2C), UMR 5221 CNRS-Universit\'{e} de Montpellier, F- 34095 Montpellier, France}

\author{Mauro Antezza}
\affiliation{Laboratoire Charles Coulomb (L2C), UMR 5221 CNRS-Universit\'{e} de Montpellier, F- 34095 Montpellier, France}
\affiliation{Institut Universitaire de France, 103 Boulevard Saint-Michel, F-75005 Paris, France}

\newcommand{\ket}[1]{\displaystyle{|#1\rangle}}
\newcommand{\bra}[1]{\displaystyle{\langle #1|}}

\date{\today}

\begin{abstract}
We study energy-transport efficiency in light-harvesting planar and 3D complexes of two-level atomic quantum systems, embedded in a common thermal blackbody radiation.
We show that the collective non-local dissipation induced by the thermal bath plays a fundamental role in energy transport. It gives rise to a dramatic enhancement of the energy-transport efficiency, which may largely overcome $100\%$. This effect, which improves the understanding of transport phenomena in experimentally relevant complexes, suggests a particularly promising mechanism for quantum energy management.
\end{abstract}

\maketitle

\section{Introduction}
Transport of energy is a phenomenon of the utmost importance in several domains of physics. Recent advancements in manipulations of systems approaching quantum scales have further triggered research on transport between single quantum entities \cite{Li2012,Michel2005,Wichterich2007}. This interest has a two-fold reason: first, the understanding of energy fluxes is necessary for controlling micro- and nanoscale devices \cite{Li2012,Narayanaswamy2003,Messina2012}; second, purely quantum effects allow for energy management not achievable in classical scenarios \cite{Plenio2013,Biehs2013,Leggio2015}. Probably the most celebrated of these effects is the role of quantum coherence in photosynthetic efficiency \cite{Ishizaki2012,Wu2013}. Stimulated by these results, quantum excitation transport in networks of two-level systems (TLSs) has recently received an enormous deal of attention \cite{Fischer2013,Feist2014,Schachenmayer2014,Celardo2014}.
In this context, the interaction of TLSs with various types of environments has been largely investigated \cite{Feist2014,Plenio2010njp, Mohseni2008}.
In particular, the effect of thermal blackbody radiation has been studied using phenomenological models only including local noise acting on each atom \cite{Plenio2010njp, Mohseni2008, Wu2010}, which typically deteriorates transport as temperature increases. This approach, neglecting non-local effects induced by the radiation, led the attention mostly on low-temperature transport, where non-locality is indeed marginal.

In this Letter, through a microscopic model, we explore excitation transport along TLSs in a common thermal blackbody radiation at arbitrary temperature, fully taking into account radiation-induced non-local effects. We show that the collective non-local dissipation due to the correlations of the thermal radiation may have a dramatic effect on transport: unexpectedly, transport efficiency may \emph{increase} with temperature up to values far beyond $100\%$.

\section{Physical system}
\begin{figure}[h!]
\begin{center}
\includegraphics[width=240pt]{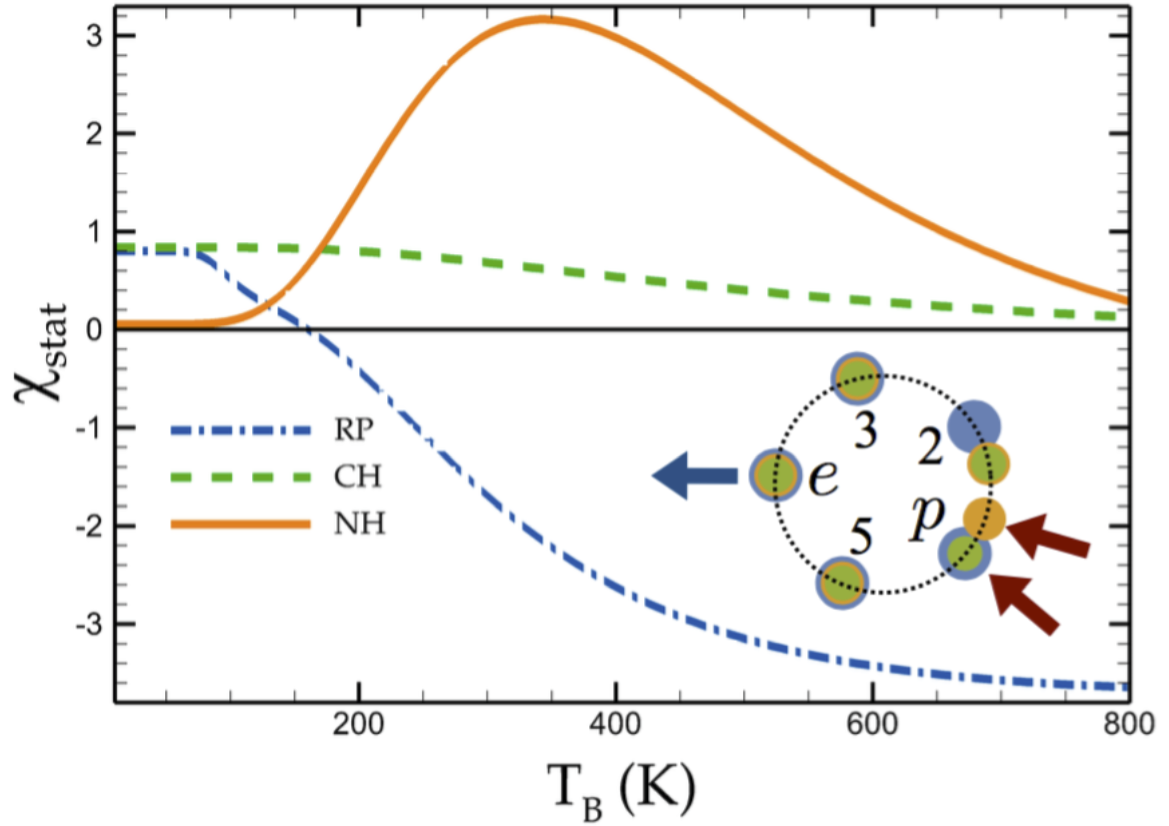}
\end{center}
\caption{Stationary transport efficiency $\chi_{\mathrm{stat}}$ versus $T_B$ for three planar configurations of 5 atoms on a ring. Atoms are either placed on the vertices of a regular pentagon (RP, blue points and dot-dashed line), one of them is displaced by an angle $\theta_2=-0.37\,\text{rad}$ (clockwise-hopping CH, green points and dashed line), or two of them are displaced in a symmetric way with $\theta_p=-\theta_2=0.37\,\text{rad}$ (no hopping NH, orange points and solid line).}
\label{2d}
\end{figure}

We consider $N$ identical two-level quantum emitters (hereby referred to as \emph{atoms}) of frequency $\omega_a$ with ground (excited) state $|0_i\rangle$ ($|1_i\rangle$), embedded in a blackbody radiation at temperature $T_B$. They couple to the electromagnetic field through their dipole moment operators, having ground-excited matrix elements all equal to $\mathbf{d}$. Without external perturbations, the atomic system equilibrates at $T_B$. Suppose now that two of these atoms are externally perturbed by pumping excitations into one and extracting energy from another (see sketch in Fig. \ref{2d}).

In the weak-coupling limit, the dynamics of the atomic density matrix $\rho$ is described by the Markovian master equation (ME) \cite{QuantInter}
\begin{equation}\label{ME}
\dot{\rho}=-\frac{\text{i}}{\hbar}\big[H_{\mathrm{tot}},\rho\big]+D_{\mathrm{loc}}\rho+ D_{\mathrm{nl}}\rho+D_{\mathrm{in}}\rho+D_{\mathrm{out}}\rho,
\end{equation}
where $H_{\mathrm{tot}}=H_a+\sum_{i\neq j}H_I^{(ij)}$, $H_a=\sum_iH_i=\hbar\omega_a\sum_i\sigma_i^+\sigma_i^-$ is the atomic free Hamiltonian and $H_I^{(ij)}=\hbar\Lambda_{ij}\sigma^-_i\sigma^+_j$ the field-induced atomic dipolar coupling, with $\sigma_i^{+(-)}$ the raising (lowering) operator of atom $i$.
\begin{equation}\label{local}
\begin{split}
D_{\mathrm{loc}}\rho&=\sum_i\bigg\{n\gamma_0\Big(\sigma^+_i\rho\sigma^-_i-\frac{1}{2}\{\sigma^-_i\sigma^+_i, \rho\}\Big)\\
&+(1+n)\gamma_0\Big(\sigma^-_i\rho\sigma^+_i-\frac{1}{2}\{\sigma^+_i\sigma^-_i, \rho\}\Big)\bigg\}
\end{split}
\end{equation}
describes standard thermal dissipation and $\gamma_0=d^2\omega_a^3/(3\hbar\pi\varepsilon_0 c^3)$.
The photon number $n=n(T_B,\omega_a)=1/[\exp(\hbar \omega_a/k_{\rm B} T_B)-1\big]$, encompasses the fundamental dependence of atomic dynamics on $T_B$. Moreover
\begin{eqnarray}\label{nonlocal}
D_{\mathrm{nl}}\rho&=&\sum_{i< j}D_{\mathrm{nl}}^{(ij)}\rho=\sum_{i< j}\bigg\{n\gamma_{ij}\Big(\sigma^+_i\rho\sigma^-_j-\frac{1}{2}\{\sigma^-_j\sigma^+_i, \rho\}\Big)\nonumber\\
&+&(1+n)\gamma_{ij}\Big(\sigma^-_i\rho\sigma^+_j-\frac{1}{2}\{\sigma^+_j\sigma^-_i, \rho\}\Big)+\text{h.c.}\bigg\}
\end{eqnarray}
describes two-body non-local dissipation, characterising energy exchanges between the field and any two-atom subsystems. These terms, neglected in phenomenological descriptions of transport based on atom-field coupling \cite{Mohseni2008, Wu2010, Manzano2013}, are a natural consequence of non-local field self-correlations in a microscopic derivation of the ME \cite{QuantInter,Bellomo2013}. Their effects are a manifestation of the quantumness of atom-field coupling: they have indeed been studied with respect to bipartite entanglement of spins in a common environment \cite{Ficek2002, Luo2007, Mancini2013}.
The rates involved in the ME \eqref{ME} are analytical functions of interatomic distances, atomic dipoles and atomic frequency \cite{QuantInter}.
In the case of identical and real dipoles $\mathbf{d}$ considered here, they result
\begin{equation}
\gamma_{ij}=\gamma_0\sum_{n=1,2,3}\Big([\hat{\mathbf{d}}]_n\Big)^2\alpha_{ij}^{(n)},
\end{equation}
where $\hat{\mathbf{d}}=\mathbf{d}/|\mathbf{d}|$, $[\hat{\mathbf{d}}]_1$ is the component along the line joining the two atoms and $[\hat{\mathbf{d}}]_{(2,3)}$ the components along directions perpendicular to each other and to the line joining the atoms.
Being $\mathbf{r}_{ij}=\mathbf{r}_j-\mathbf{r}_i$ the spatial separation of two atoms, $\tilde{r}_{ij}=\frac{\omega_a}{c}|\mathbf{r}_{ij}|$, $\hat{\mathbf{r}}_{ij}=\frac{\mathbf{r}_{ij}}{|\mathbf{r}_{ij}|}$, coefficients $\alpha_{ij}^{(n)}$ are
\begin{eqnarray}
\alpha_{ij}^{(1)}&=&\frac{3}{\tilde{r}_{ij}^3}\Big(\sin{\tilde{r}_{ij}}-\tilde{r}_{ij}\cos{\tilde{r}_{ij}}\Big),\\
\alpha_{ij}^{(2)}&=&\alpha_{ij}^{(3)}=\frac{3}{2\tilde{r}_{ij}^3}\Big(\tilde{r}_{ij} \cos{\tilde{r}_{ij}}+(\tilde{r}_{ij}^2-1)\sin{\tilde{r}_{ij}}\Big).
\end{eqnarray}
Finally, the dipole-dipole interaction strength is
\begin{equation}\begin{split}
\Lambda_{ij}&=-\frac{3}{4}\gamma_0\Big[2(\hat{\mathbf{d}}\cdot\hat{\mathbf{r}}_{ij})^2f(\tilde{r}_{ij})+\Big(1-(\hat{\mathbf{d}}\cdot\hat{\mathbf{r}}_{ij})^2\Big)g(\tilde{r}_{ij})\Big],\label{lambda}\\
f(x)&=\frac{\cos{x}+x\sin{x}}{x^3},\quad g(x)=\frac{(x^2-1)\cos{x-x\sin{x}}}{x^3}.
\end{split}\end{equation}
Incoherent pumping on the atom labeled as $p$ is accounted for by the term
\begin{equation}\label{Din}
D_{\mathrm{in}}\rho=\gamma_{\mathrm{in}}\left(\sigma^+_p\rho\sigma^-_p-\frac{1}{2}\{\sigma^-_p\sigma^+_p,\rho\}\right).
\end{equation}
Another external operator (the sink) extracts energy locally from the atom labeled as $e$, through the process
\begin{equation}\label{Dout}
D_{\mathrm{out}}\rho=\gamma_{\mathrm{out}}\left(\sigma^-_e\rho\sigma^+_e-\frac{1}{2}\{\sigma^+_e\sigma^-_e,\rho\}\right).
\end{equation}

In general, a Markovian dissipative term $D$ produces an energy flux $\mathrm{Tr}\big(H_S D\rho\big)$, $H_S$ being the free Hamiltonian of the open system \cite{BreuerBook}. According to this definition, energy flowing out of an open system is negative.
The rates of energy pumping $P(t)$ and extraction $E(t)$ are given by
\begin{eqnarray}
P(t)&=&\frac{\mathrm{d}}{\mathrm{d}t}U_P(t)=\mathrm{Tr}\big(H_a D_{\mathrm{in}}\rho(t)\big),\\
E(t)&=&\frac{\mathrm{d}}{\mathrm{d}t}U_E(t)=-\mathrm{Tr}\big(H_a D_{\mathrm{out}}\rho(t)\big),
\end{eqnarray}
$U_{P(E)}(t)$ being the total energy pumped in (extracted from) the atomic system up to time $t$. By construction $P(t),E(t)\geq0$, and depend on $T_B$ through the atomic state $\rho(t)$.
Due to thermal excitations, energy is extracted from $e$ also in the absence of pumping ($\gamma_{\mathrm{in}}=0$): the sink, through atom $e$, absorbs thermal photons from the field at a rate $E_0$. To evaluate how effective the transport of excitations from $p$ to $e$ is, one can analyse the contribution to extracted flux due to pumping only, as expressed by the transport efficiency
\begin{equation}\label{chi}
\chi(t)=\frac{E(t)-E_0(t)}{P(t)}.
\end{equation}
When $\chi(t)>1$, $E>E_0+P$, i.e., the extracted excitations are more than the sum of the thermal and the pumped ones: as will be shown in the following, efficiency will largely overcome $100\%$  thanks to the combination of thermal noise and non-local effects. The definition of $\chi$ does not include energy fluxes between atoms and the blackbody radiation and, as such, $\chi$ is not upper-limited by $1$. It is therefore not a thermodynamic efficiency but rather characterises how efficiently pumping affects extraction, as these are the only two quantities on which an external control is possible.

At stationarity, all the quantities involved in Eq. \eqref{chi} become time-independent, hence both $U_P$ and $U_E$ diverge linearly in $t$ and
\begin{equation}\label{chistat}
\chi_{\mathrm{stat}}=\frac{U_E^{\mathrm{stat}}-U^{\mathrm{stat}}_{E_0}}{U_P^{\mathrm{stat}}}
\end{equation}
is the fraction of total pumped energy one recovers at the extraction site.

Following Eq. \eqref{ME}, atoms have two ways of affecting each other's internal energy: coherent interaction $H_I$ and collective dissipative exchanges $D_{ij}^{\mathrm{nl}}$ with the environment. The associated energy fluxes are $\Q_{ij}^{\mathrm{hop}}(t)=\frac{1}{\text{i}\hbar}\mathrm{Tr}\big(H_j [H_I^{(ij)},\rho(t)] \big)$ (flux of energy \textit{from} i \textit{to} j) and $\Q_{ij}^{\mathrm{nl}}(t)=\mathrm{Tr}\big(H_j D_{\mathrm{nl}}^{(ij)}\rho(t)\big)$ (energy flowing non-locally from the field to each of the atoms $i,j$, i.e., two photons are emitted or absorbed simultaneously by the two atoms). Explicit evaluation gives
\begin{eqnarray}
\Q_{ij}^{\mathrm{nl}}(t)&=&-\frac{\hbar}{2}\omega_a\gamma_{ij}\langle \sigma_i^+\sigma_j^-+(\sigma_i^+\sigma_j^-)^{\dag} \rangle,\label{Qnl}\\
\Q_{ij}^{\mathrm{hop}}(t)&=&\text{i}\hbar\omega_a\Lambda_{ij}\langle \sigma_i^+\sigma_j^--(\sigma_i^+\sigma_j^-)^{\dag} \rangle.\label{Qhop}
\end{eqnarray}
The correlators of emission and absorption processes satisfy $\langle \sigma_i^+\sigma_j^-+(\sigma_i^+\sigma_j^-)^{\dag} \rangle=2\mathrm{Re}(c^{ij}(t))$ and $\langle \sigma_i^+\sigma_j^--(\sigma_i^+\sigma_j^-)^{\dag} \rangle=2\text{i}\mathrm{Im}(c^{ij}(t))$, $c^{ij}(t)$ being the quantum coherence between the states $|0_i1_j\rangle$ and $|1_i0_j\rangle$ in $\rho_{ij}=\mathrm{Tr}_{a\neq i,j}\rho$. We remark that dissipation plays here an active role in transporting excitations coherently [Eq. \eqref{Qnl}], contrarily to previous models where its effect mostly consisted in localising excitations by suppressing coherences.

\section{Transport in planar configurations}
We apply now this theory to transport through five-atom rings of radius $r=0.4\,\mu$m, as in Fig. \ref{2d}. Planar ring configurations have been studied as a simpler prototype of light-harvesting physics \cite{Celardo2014}. The atomic frequency is $\omega_a=10^{14}\,\text{rad/s}$ ($\hbar\omega_a\simeq6.6\cdot 10^{-2}\,\text{eV}$) and dipoles lie along the axis perpendicular to the ring. These parameters are in accordance with some typical values found in, e.g., quantum dots \cite{Barreiro2012}, hyperfine structure of real atoms and molecular vibrational states. Moreover, although not usually met in biological light-harvesting complexes \cite{QuantEffBio}, these parameters will permit us to highlight more clearly the general importance of the non-local dissipation in transport phenomena. The pumping rate is $\gamma_{\mathrm{in}}=10^{-3}\,\gamma_0$, to simulate low-intensity light impinging on the system, while the extraction rate $\gamma_{\mathrm{out}}=10^2\,\gamma_0$, in accordance with typical sink rates in light harvesting complexes \cite{QuantEffBio,Plenio2008}. Thanks to this choice, all the term in the ME are proportional to $\gamma_0$. In what follows we will therefore work in units of $\gamma_0$.
Pumping and extraction are exerted as shown in Fig. \ref{2d}. All the results reported hereafter are obtained using the open-source QuTiP package \cite{QuTiP}.

Three different geometrical dispositions are considered. First, atoms are placed on the vertices of a regular pentagon (RP), as shown in Fig. \ref{2d} (blue points). In this case two paths, clock- and counterclockwise, exist for excitations to travel to atom $e$. If $T_B$ is high enough, it is known that field-induced effects can create multipartite correlations in symmetric atomic configurations \cite{Bellomo2015}: excitations get delocalised over the whole atomic system and transport efficiency is strongly reduced.
Fig. \ref{2d} shows that transport is indeed rapidly killed by thermal noise.

The second configuration is obtained through a displacement of atom 2 from its regular position of an angle $\theta_2=-0.37\,\text{rad}$ (green dots in Fig. \ref{2d}). We refer to this configuration as clockwise hopping (CH): the fast decay of $\Lambda_{ij}$ with the interatomic distance $r_{ij}$ induces a bias between the two hopping directions. Excitations are mostly transmitted along the path $p\rightarrow5\rightarrow e$ with fewer destructive interference: $\chi_{\mathrm{stat}}$ for this configuration is depicted in Fig. \ref{2d} (dashed green line). In accordance with hopping physics, local noise disturbs excitation transmission, continuously reducing its efficiency with increasing $T_B$.

Finally, also the second pathway is killed by displacing the atom $p$ closer to atom $2$, such that a symmetric configuration is obtained (orange points in Fig. \ref{2d}). A natural expectation is that transport becomes hardly possible due to the vanishing hopping amplitude (no-hopping configuration NH). Nonetheless, $\chi_{\mathrm{stat}}$ shows a pronounced maximum in $T_B$, at efficiencies higher than $300\%$: remarkably when hopping is killed, the increase in extracted excitations becomes larger than pumping. There being no hopping, such an enormous gain of excitations origins from $D_{ij}^{\mathrm{nl}}$, by which atoms can affect each other's internal energy through collective field-induced dissipation.

\begin{figure}[h!]
\begin{center}
\includegraphics[width=240pt]{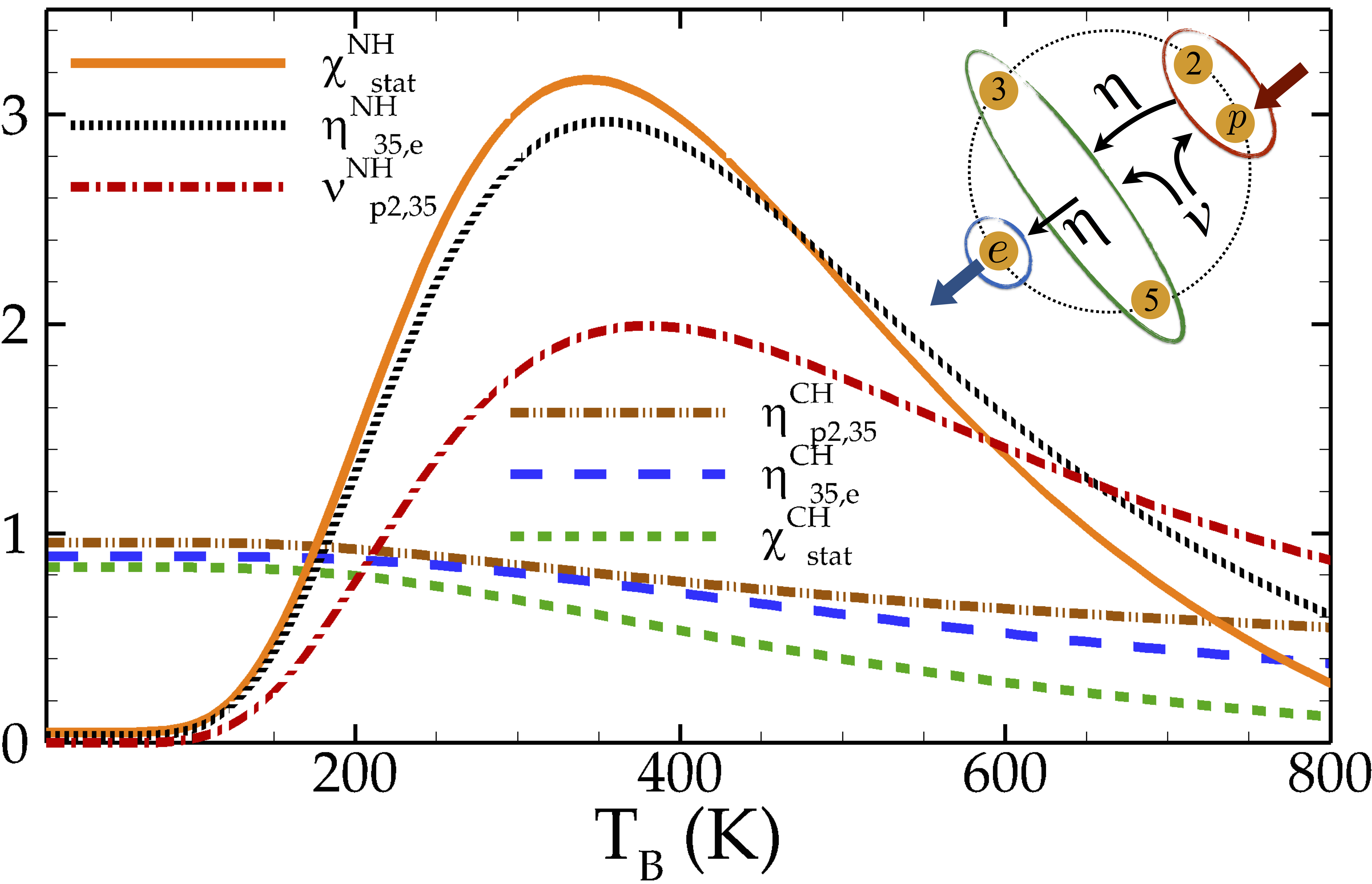}
\end{center}
\caption{$\chi_{\mathrm{stat}}^{\mathrm{NH}}$ (orange solid line), $\chi_{\mathrm{stat}}^{\mathrm{CH}}$ (green dashed line) and the principal contributions to all stages of the main path. The first stage $\{p,2\}\rightarrow\{3,5\}$ in CH and NH configurations happens through different physical effects: hopping $\eta$ for CH, non-local dissipation $\nu$ for NH.}
\label{nlhop}
\end{figure}

To better understand this mechanism, the atomic system can be divided into three subsets: the pair $\{p,2\}$ where excitations arrive, the pair $\{3,5\}$ which transmits these excitations and the monomer $\{e\}$ in contact with the sink, as shown in the inset of Fig. \ref{nlhop}.
Only a small fraction of excitations jumps directly from $p$ to $e$, the largest part travelling along the main path $\{p,2\}\rightarrow\{3,5\}\rightarrow\{e\}$.
Since local heat fluxes can not affect transport due to the Markovian structure of the environment, the only possible mechanisms responsible for excitation transmission between atoms are hopping and non-local dissipation. To analyse them, we introduce
\begin{eqnarray}
\eta_{ij,ab}&=&\sum_{\substack{\alpha=a,b\\ \kappa=i,j}}\frac{\Q_{\kappa\alpha}^{\mathrm{hop}}(\infty)\big|_{\gamma_{\mathrm{in}}}-\Q_{\kappa\alpha}^{\mathrm{hop}}(\infty)\big|_{0}}{P},\\
\nu_{ij,ab}&=&\sum_{\substack{\alpha=a,b\\ \kappa=i,j}}\frac{\Q_{\kappa\alpha}^{\mathrm{nl}}(\infty)\big|_{\gamma_{\mathrm{in}}}-\Q_{\kappa\alpha}^{\mathrm{nl}}(\infty)\big|_{0}}{P},
\end{eqnarray}
representing the fraction of pumped excitations travelling to the subset $\{a,b\}$ through either hopping ($\eta$) from or non-local dissipation ($\nu$) with each atom of the subset $\{i,j\}$. Fig. \ref{nlhop} shows, for both NH and CH configurations, the quantities $\chi_{\mathrm{stat}}$ and $\eta_{35,e}$ versus $T_B$. It is clear that the mechanism responsible for the last stage $\{3,5\}\rightarrow\{e\}$ of the main path is hopping: the curves $\chi_{\mathrm{stat}}^{\mathrm{NH}(\mathrm{CH})}$ (solid orange and dotted green lines) and $\eta_{35,e}^{\mathrm{NH}(\mathrm{CH})}$ (dotted black and dashed blue lines) are remarkably similar, showing that almost all the energy extracted from $e$ arrives there through hopping from atoms 3 and 5.

A striking difference appears however in the first stage $\{p,2\}\rightarrow\{3,5\}$: in CH case, it is again mostly due to hopping from pair $\{p,2\}$ to pair $\{3,5\}$, as shown by $\eta_{p2,35}^{\mathrm{CH}}$ (brown dot-dot-dashed line). Non-local transport $\nu_{p2,35}^{\mathrm{CH}}$ (not shown here) is indeed negligible. On the other hand, the stage $\{p,2\}\rightarrow\{3,5\}$ in NH setup has a large contribution from collective non-local dissipation as highlighted by the non-local energy exchanges $\nu_{p2,35}^{\mathrm{NH}}$ (red dot-dashed line): the pumping of excitations on $p$ triggers the electromagnetic field to inject photons into both atomic pairs.
Note that, as a consequence, excitations received by $\{3,5\}$ can exceed the ones pumped into $p$, the difference being supplied by the field: $\nu_{p2,35}^{\mathrm{NH}}$ accounts almost exactly for the additional parts of excitations travelling to $e$. Energy arriving on $\{3,5\}$ completes then its path by hopping to $e$.
\begin{figure}[h!]
\begin{center}
\includegraphics[width=240pt]{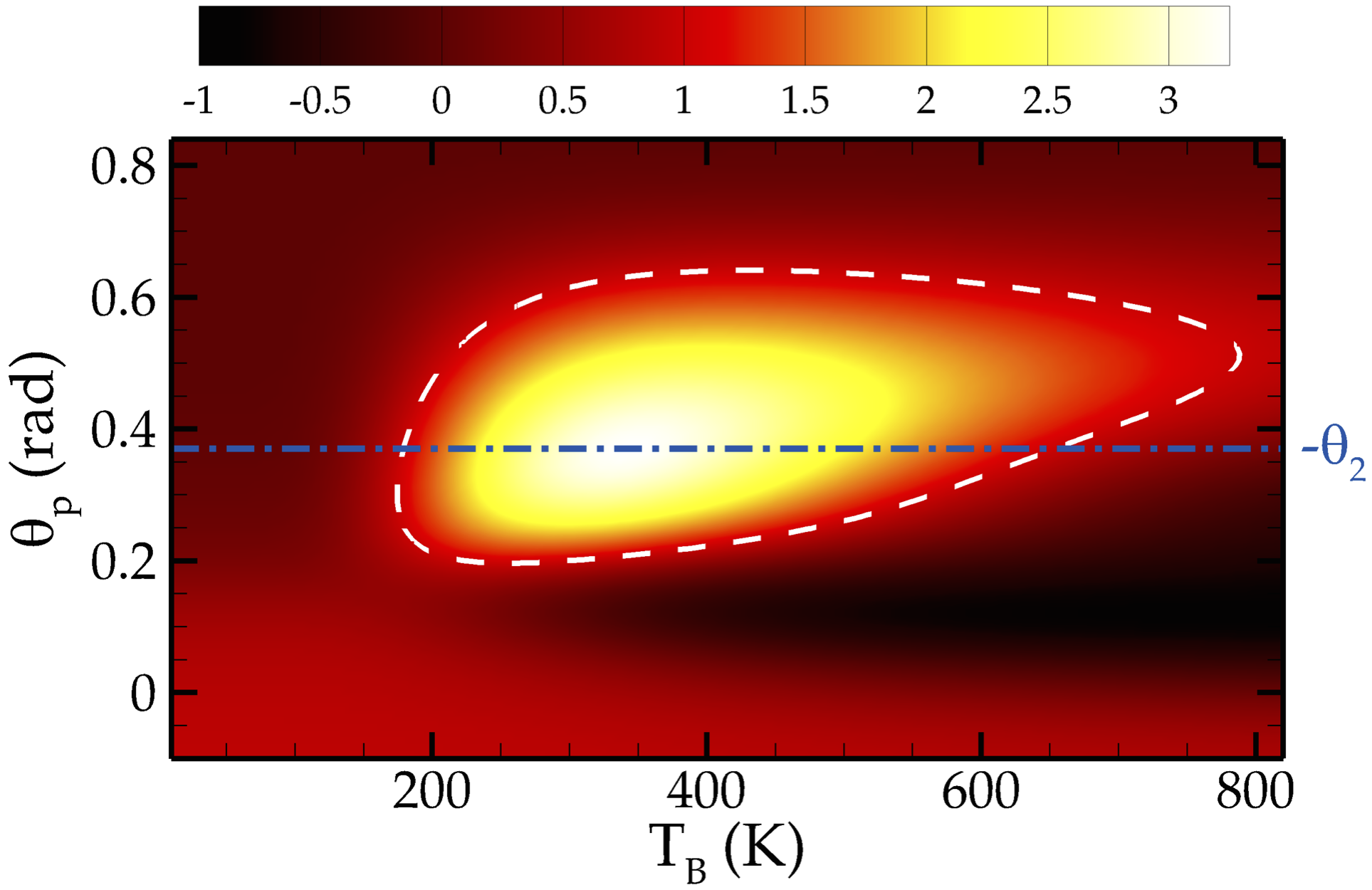}
\end{center}
\caption{$\chi_{\mathrm{stat}}$ versus $T_B$ and angular displacement $\theta_p$ of $p$, with $\theta_2=-0.37\,\text{rad}$. The dashed white line is the contour $\chi_{\mathrm{stat}}=1$. The horizontal dot-dashed line represents the NH configuration ($\theta_p=-\theta_2=0.37\,\text{rad}$), where transport efficiency is maximised.}
\label{angl}
\end{figure}
Following Eq. \eqref{Qnl}, non-local energy flux is proportional to the real part of the quantum coherence between two atoms. Transport in NH configuration is thus a dissipation-induced coherent effect.
The importance of the symmetric positions of atoms $p$ and $2$ in creating a pair is proved by the dependence of $\chi_{\mathrm{stat}}$ on the angular displacement $\theta_p$ of $p$, reported in Fig. \ref{angl}. There, $\theta_p=0$ corresponds to CH configuration. Note that the role of symmetry on quantum correlations and quantum transport was already pointed out in different contexts \cite{Bellomo2015,Lloyd2010}.

\begin{figure*}[t!]
\includegraphics[width=500pt]{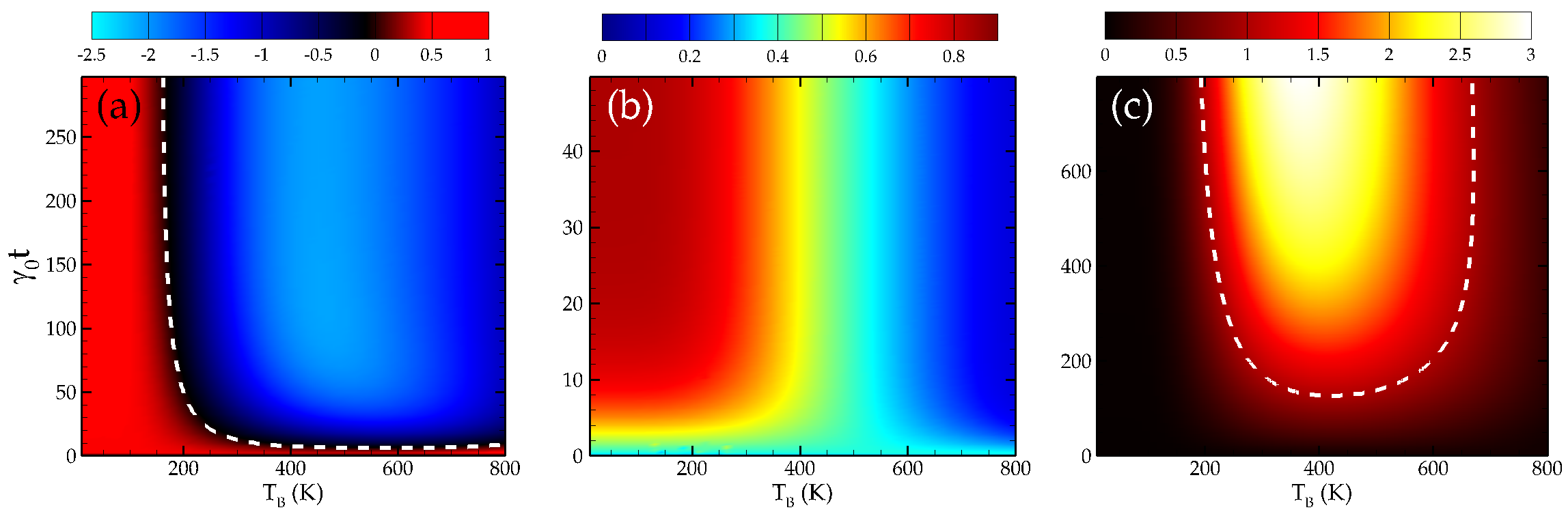}
\caption{Density plot of $\chi(t)$ for (a): RP, (b): CH, (c): NH, versus time and $T_B$. The white dashed line in (a) represents the contour $\chi(t)=0$, while in (c) it represents the contour $\chi(t)=1$. Neither of the two contours appear in CH setup. Dynamics starts from $|\psi_0\rangle=\bigotimes_{i=1}^5|0_i\rangle$.}
\label{chit}
\end{figure*}

The reason why these non-local effects do not contribute when hopping is significant is twofold: first, when hopping is active there can be no isolated subset of atoms where pumping is exerted, while such subset is present in NH configuration (pair $\{p,2\}$). This subset, having a much larger excited-state population than the rest of the system due to pumping on it, drives the collective atomic state further away from a purely thermal state, thus allowing the establishing of steady quantum coherence: we develop in the next section a simple toy model able to grasp this mechanism. The second reason comes from the timescales of different dynamical processes: Fig. \ref{chit} shows the time- and $T_B$-dependence of $\chi(t)$ in RP [panel (a)], CH [panel (b)] and NH [panel (c)] configurations with initial state $|\psi_0\rangle=\bigotimes_{i=1}^5|0_i\rangle$.

Compare panels (b) and (c): transport through hopping (b) happens on much shorter timescales than the non-local one, as expectable since $\Lambda_{ij}\gg\gamma_{ij}$. As long as hopping is active, non-locality cannot play any substantial role: by the time non-locality becomes effective, excitations have already travelled through hopping. Excitations reach $e$ in $\gamma_0 t<1$ in CH setup and stationarity is attained for $\gamma_0 t<10$. On the other hand, $\chi^{\mathrm{NH}}(t)$ becomes non-negligible in $\gamma_0 t\gtrsim 10$, while stationarity is reached for $\gamma_0 t\gtrsim 600$. Nonetheless, at suitable temperatures, $\gamma_0 t\simeq 100$ is enough for transport efficiency in NH configurations to go beyond $1$ (white dashed line in panel (c)), fully entering the non-local amplification regime.

\section{Thermally-induced coherence amplification: a toy model}\label{toymodel}
The pairs $\{p,2\}$ and $\{3,5\}$ in NH configurations are almost not connected by hopping. Excitations pumped in $\{p,2\}$ can thus hardly leave the pair, and the excited state populations of atoms $p$ and 2 grow remarkably with respect to those in $\{3,5\}$. The effect is thus to create a gap in populations between two branches of the atomic system. This means that the system itself is no longer in the vicinity of thermal equilibrium and can thus sustain quantum coherence between its subparts. To understand this on a more quantitative basis, consider a toy model of two atoms, having the same ME as Eq. \eqref{ME} but with $\Lambda_{ij}=\gamma_{\mathrm{in}}=\gamma_{\mathrm{out}}=0$. Moreover, a population gap is imposed by two different blackbody reservoirs, one for each atom, such that $n$ in Eqs. \eqref{local}-\eqref{nonlocal} is different for the two atoms.
In particular, one atom is connected to a hot reservoir with $n=n_h$ (simulating the effect of pumping), while the other one is connected to a reservoir with $n=n_B<n_h$. Non-local effects are also assumed to happen at $n_B$. The atoms coherence $c$ in the decoupled basis is always real and depends on $n_h$ and $n_B$ as
\begin{equation}
c(n_B,n_h)=\frac{\gamma_0 \gamma_{\mathrm{nl}}(1+n_h+n_B)(n_B-n_h)}{2F(n_B,n_h)},
\end{equation}
where
\begin{equation}\begin{split}
F(n_B,n_h)&=\gamma_0^2(1+2n_B)(1+2n_h)(1+n_B+n_h)^2\\
&\,-\gamma_{\mathrm{nl}}^2\big[1+n_B(7+13n_B+8n_B^2)+n_h\\
&\,+4n_hn_B(3+6n_B+4n_B^2)-n_h^2\big].
\end{split}\end{equation}
Under the natural condition $\gamma_0>\gamma_{\mathrm{nl}}$, $|c(n_B,n_h)|$ shows a maximum as a function of $n_h$ at $n_h=n_h^*$.
Despite its simplicity, then, this toy model gives already important hints to better understand the physics of non-local transport: driving a collective atomic state far from its Gibbs form by imposing a population gap between two parts generates real quantum coherence between them, with a sharp peak when an optimal gap is achieved. This, in turn, means a sharp peak in the non-local energy flux involving the two subparts.
The symmetric displacement characterising NH configurations produces exactly such an optimal value of population gap.

\section{Transport of excitations in FMO-inspired configurations}
We consider next a 3D setup inspired by the celebrated Fenna-Matthews-Olson (FMO) complex \cite{QuantEffBio}, a pigment-protein complex used for photosynthesis. Light is collected and transmitted by 7 molecular complexes (schematically seen as TLSs) to a sink, where it is used for biological functions. Nowadays, its high light-transport efficiency is believed to have quantum origins \cite{Ishizaki2012,QuantEffBio}. Its transport properties, as a function of temperature of a photonic bath, have been studied in \cite{Mohseni2008}, but non-local dissipation has there been neglected.

\begin{figure}[t!]
\begin{center}
\includegraphics[width=240pt]{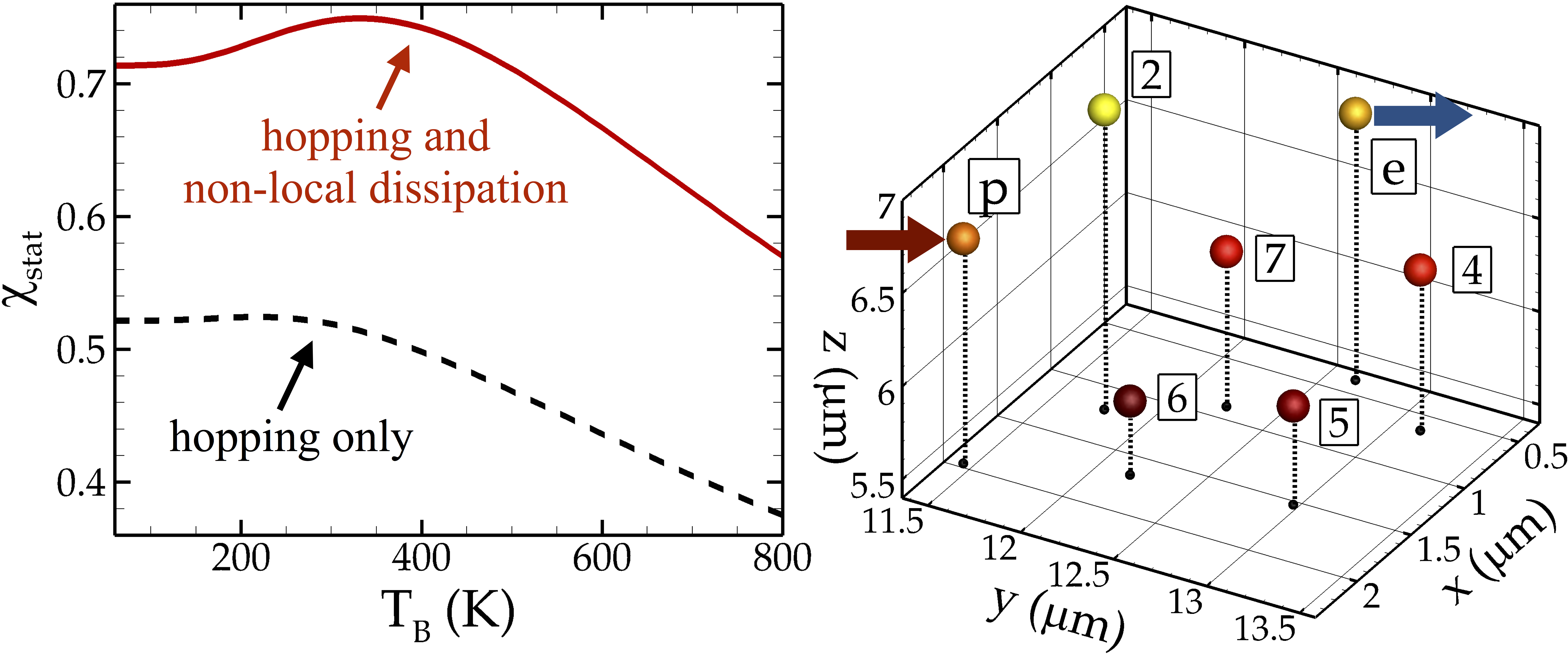}
\end{center}
\caption{$\chi_{\mathrm{stat}}$ versus $T_B$ in FMO geometry, with site-site distances scaled with respect to the FMO complex by a factor $0.7\cdot10^{3}$. The black dashed line represents the case $\gamma_{ij}=0$, the solid red line is the efficiency obtained when $\gamma_{ij}\neq0$.}
\label{fmo}
\end{figure}

The spatial configuration of the 7-site system we study is shown in Fig. \ref{fmo}, obtained from the FMO site configuration (given for instance in \cite{Lloyd2014}) by rescaling each site-site distance by a factor $700$. Typical lifetimes of excitations in FMO complexes are $\sim1\,\text{ns}$ \cite{Plenio2008}, while extraction time can be of the order of $0.25\,\text{ps}$: we set $\gamma_{\mathrm{out}}/\gamma_0=2.5\cdot10^2$. The ratio $\gamma_{\mathrm{in}}/\gamma_0=10^{-3}$ is compatible with the low-intensity light impinging on the molecule, having verified this value to be low enough for the $\gamma_{\mathrm{in}}$-dependence of $\chi_{\mathrm{stat}}$ to be flat. There being no preferential direction is this system, we choose the dipoles to lie along the $x$-axis of Fig. \ref{fmo}.

It is important to stress that, by this configuration, we do not intend to analyse the exact physics of transport in FMO complexes where transition frequencies, dipole orientations, intersite distances and environment are much different from the ones considered here \cite{Mohseni2008,QuantEffBio,Plenio2008,Lloyd2014}. Nevertheless, since the relative weight of non-local effects over hopping is ultimately due to configurations of sites in a network, uncovering their role in a geometry analogous to the FMO might shed light on equivalent effect during photosynthesis and suggest how to enhance transport performances in artificial realisations. The FMO geometry shows indeed two sites ($p$ and $2$ in Fig. \ref{fmo}), relatively closer to each other than to the rest of the complex.

Plots in Fig. \ref{fmo} show $\chi_{\mathrm{stat}}$ versus $T_B$ in the absence (dashed black line) and in the presence (solid red line) of $\gamma_{ij}$: the presence of non-local effects produces two interesting features: first, $\chi_{\mathrm{stat}}$ is always higher if $\gamma_{ij}\neq0$. Second, it introduces a non-monotonic dependence on $T_B$ very similar to the one shown in Fig. \ref{2d}, suggesting an important contribution of non-local energy transfer.

For completeness, we have also simulated the exact FMO complex, where absorption happens mostly around $\lambda=800\,\text{nm}$ ($\omega_a=2.35\cdot10^{15}\,\text{rad/s}$) \cite{QuantEffBio}, using the exact intersite distances and dipole orientations \cite{Lloyd2014}. For this high frequency, $n(T_B,\omega_a)$ does not significantly change around room temperature and the $T_B$-dependence of $\chi_{\mathrm{stat}}$ is flat. Nonetheless, non-local effects still play a significant role: $\chi_{\mathrm{stat}}\simeq 0.8$ when $\gamma_{ij}=0$, while $\chi_{\mathrm{stat}}\simeq 0.88$ if $\gamma_{ij}\neq0$, increasing transport efficiency of $\sim 10\%$.

\section{Conclusions}
We showed that the energy transport in atomic quantum systems can be considerably affected by embedding them in an electromagnetic thermal bath.
Non-local effects induced by the thermal field, neglected in previous investigations, are shown to play an active role in coherent excitation transport and to boost its efficiency, which grows with temperature, up to values largely overcoming $100\%$ in planar configurations. This mechanism originates from non-local self-correlations of electromagnetic radiation, which induce collective dissipation in atoms: when they are brought out of equilibrium by local pumping and extraction, thermally-activated collective effects inject a great amount of excitations in the system and drastically increase energy extraction from it.

These results suggest to exploit the intrinsic non-locality of thermal noise as a resource to improve quantum energy transport, with direct applications to condensed-matter systems, to nanoscale energy management, and possibly being of great relevance also for light-harvesting

\begin{acknowledgements}
We acknowledge financial support from the Julian Schwinger Foundation.
\end{acknowledgements}

\end{document}